\documentclass[10pt, conference]{IEEEtran}
\usepackage{amsfonts}
\usepackage{fancyhdr}

\def\BibTeX{{\rm B\kern-.05em{\sc i\kern-.025em b}\kern-.08em
 T\kern-.1667em\lower.7ex\hbox{E}\kern-.125emX}}

\usepackage{booktabs}

\usepackage[printonlyused,nolist]{acronym} 

\usepackage{pifont}
\newcommand{\cmark}{\ding{51}}%
\newcommand{\xmark}{\ding{55}}%

\newcommand{\drop}[1]{\textcolor{red}{#1}}
\renewcommand{\drop}[1]{}

\usepackage[linesnumbered,ruled]{algorithm2e}
\usepackage{circledsteps}
\pgfkeys{/csteps/inner color=white}
\pgfkeys{/csteps/outer color=none}
\pgfkeys{/csteps/fill color=black}
\newdimen\arrayruleHwidth
\newcommand{\blue}[1]{\textcolor{black}{#1}}
\newcommand{\bluee}[1]{\textcolor{black}{#1}}
\newcommand{\rede}[1]{\textcolor{black}{#1}}
\usepackage{balance}
\usepackage{placeins}
\usepackage{upgreek}
\usepackage{hyperref}

\usepackage{multirow}
\usepackage{tikz}

\newcommand{\DrawBar}[1]{%
  \begin{tikzpicture}
\fill[color=black]   (0.0 , 0.0) rectangle (#1/21*40ex , 1.5ex );
\fill[color=white] (#1/21*40ex  , 0.0) rectangle (40ex, 1.5ex);
  \end{tikzpicture}%
}

\usepackage{comment}

\usepackage{algorithmic}
\usepackage{graphicx}
\usepackage{textcomp}
\usepackage{xcolor}
\def\BibTeX{{\rm B\kern-.05em{\sc i\kern-.025em b}\kern-.08em
 T\kern-.1667em\lower.7ex\hbox{E}\kern-.125emX}}

\usepackage{soul}
\soulregister\cite7
\soulregister\autoref7
\soulregister\ref7
\soulregister\cref7
\soulregister\Cref7
\soulregister\ 7
\soulregister\texttt7
\soulregister\ac7
\soulregister\acp7
\soulregister\acf7
\soulregister\acl7
\soulregister\acs7
\definecolor{lightyellow}{rgb}{1.0,0.97,0.67}
\definecolor{lightgreen}{rgb}{0.5,0.99,0.6}
\definecolor{lightblue}{rgb}{0.7,0.8,0.99}
\definecolor{lightred}{rgb}{0.99,0.75,0.7}
\let\normalhl\hl
\renewcommand{\hl}{\sethlcolor{lightyellow}\normalhl}

\usepackage[printonlyused,nolist]{acronym} 

\begin{document}
\definecolor{cadmiumgreen}{rgb}{0.0, 0.42, 0.24}

\begin{acronym}
 \setlength{\itemsep}{0.2em}
 \acro{3DES}{Triple DES}
 \acro{AA}{Active Authentication}
 \acro{ADC}{Analog to Digital Converter}
 \acro{ACK}{Acknowledgement}
 \acro{AES}{advanced encryption standard}
 \acro{AID}{Application Identifier}
 \acro{ALU}{Arithmetic Logic Unit}
 \acro{ANF}{Algebraic Normal Form}
 \acro{APDU}{Application Protocol Data Unit}
 \acro{API}{Application Programming Interface}
 \acro{ASCII}{American Standard Code for Information Interchange}
 \acro{ASCA}{Algebraic Side-Channel Analysis}
 \acro{ASK}{Amplitude-Shift Keying}
 \acro{ASIC}{Application Specific Integrated Circuit}
 \acro{ATE}{Automatic Test Equipment}
 \acro{ATQA}{Answer To Request A}
 \acro{ATR}{Answer To Reset}
 \acro{ATS}{Answer To Select}
 \acro{BAC}{Basic Access Control}
 \acro{BIST}{Built-In Self-Test}
 \acro{BPSK}{Binary Phase Shift Keying}
 \acro{BSI}{Bundesamt für Sicherheit in der Informationstechnik}
 \acro{BRAM}{Block RAM}
 \acro{CBC}{Cipher Block Chaining}
 \acro{CC}{Common Criteria}
 \acro{CCA}{Canonical Correlation Analysis}
 \acro{CED}{Concurrent Error Detection}
 \acro{CHES}{Cryptographic Hardware and Embedded Systems}
 \acro{CIA}{Combined Implementation Attacks} 
 \acro{CISC}{Complex Instruction Set Computing}
 \acro{CL}{Cascade Level}
 \acro{CMRR}{Common Mode Rejection Ratio}
 \acro{CMTF}{Combined Masking in Tower Fields}
 \acro{CMOS}{Complementary Metal Oxide Semiconductor}
 \acro{COPACOBANA}{Cost-Optimized Parallel Code Breaker and Analyzer}
 \acro{COSY}{Communication Security}
 \acro{CPA}{Correlation Power Analysis}
 \acro{CPU}{Central Processing Unit}
 \acro{CRC}{Cyclic Redundancy Check}
 \acro{CRT}{Chinese Remainder Theorem}
 \acro{CSP}{cloud service provider}
 \acroplural{CSP}[CSPs]{cloud service providers}
 \acro{CTR}{Counter \acroextra{(mode of operation)}}
 \acro{CUT}{Circuit Under Test}
 \acro{DAC}{Digital-Analog Converter}
 \acro{DECT}{Digital Enhanced Cordless Telecommunications}
 \acro{DES}{data encryption standard}
 \acro{DEMA}{Differential Electro-Magnetic Analysis}
 \acro{DfT}{Design-for-Testability}
 \acro{DC}{Direct Current}
 \acro{DCM} {Digital Clock Manager}
 \acro{DFA}{differential fault analysis}
 \acro{DFIA}{Differential Fault Intensity Analysis}
 \acro{DFT}{Discrete Fourier Transform}
 \acro{DoM}{Difference of Means}
 \acro{DoS}{denial-of-service}
 \acro{DIP}{Dual Inline Package}
 \acro{DIMM}{Dual In Line Memory Modules}
 \acro{DOM}{Difference-of-Means}
 \acro{DPA}{Differential Power Analysis}
 \acro{DRAM}{Dynamic Random Access Memory}
 \acro{DRC}{design rule checking}
 \acro{DSO}{Digital Storage Oscilloscope}
 \acro{DSP}{Digital Signal Processing}
 \acro{DST}{Digital Signature Transponder}
 \acro{DTW}{Dynamic Time Warping}
 \acro{DUT}{Device Under Test}
 \acro{DVFS}{Dynamic Voltage Frequency Scaling}
 \acroplural{DUT}[DUTs]{Devices Under Test}
 \acro{DVB-T}{Digital Video Broadcasting -- Terrestrial}
 \acro{e-h}{electron-hole}
 \acro{EAC}{Extended Access Control}
 \acro{ECB}{Electronic Code Book}
 \acro{ECC}{Elliptic Curve Cryptography}
 \acro{ECDLP}{Elliptic Curve Discrete Logarithm Problem}
 \acro{ECDSA}{Elliptic Curve Digital Signature Algorithm}
 \acro{ECMA}{European Computer Manufacturers Association}
 \acro{EDC}{Error Detecting Code}
 \acro{EDE}{Encrypt-Decrypt-Encrypt \acroextra{(mode of operation)}}
 \acro{EEPROM}{Electrically Erasable Programmable Read-Only Memory}
 \acro{EM}{electromagnetic}
 \acro{EMC}{Electro-Magnetic Compatibility}
 \acro{EMSEC}{Embedded Security}
 \acro{EOC}{End Of Communication}
 \acro{EOF}{End Of File}
 \acro{ePass}{Electronic Passport}
 \acro{EPC}{Electronic Product Code}
 \acro{EPROM} {Erasable Programmable Read-Only Memory}
 \acroplural{EPROM}[EPROMs]{Erasable Programmable Read-Only Memories}
 \acro{EU}{European Union}
 \acro{FD}{Fault Dictionary}
 \acro{FDT}{Frame Delay Time}
 \acro{FF}{Flip Flop}
 \acro{FFT}{Fast Fourier Transform}
 \acro{FI}{Fault Injection}
 \acro{FIA}{Fault Injection Attack}
 \acro{FIB}{Focused Ion Beam}
 \acro{FIFO}{First In First Out \acroextra{(memory)}}
 \acro{FIMA}{Fault Intensity Map Analysis}
 \acro{FIPS}{Federal Information Processing Standard}
 \acro{FIR}{Finite Impulse Response} 
 \acro{FIT}{Faculty of Information Technology}
 \acro{FLDA}{Fisher's \ac{LDA}}
 \acro{FPGA}{Field Programmable Gate Array}
 \acro{FSA}{Fault Sensitivity Analysis}
 \acro{FSM}{Finite-State Machine}
 \acro{FSK}{Frequency Shift Keying}
 \acro{GIAnT}{Generic Implementation Analysis Toolkit}
 \acro{GND}{Ground}
 \acro{GNFS}{General Number Field Sieve}
 \acro{GUI}{Graphical User Interface}
 \acro{GPIO}{General Purpose I/O}
 \acro{GPL}{GNU General Public License}
 \acro{GPS}{Global Positioning System}
 \acro{GSM}{Global System for Mobile Communications}
 \acro{HAC}{Handbook of Applied Cryptography}
 \acro{HD}{Hamming Distance}
 \acro{HDL}{Hardware Description Language}
 \acro{HF}{High Frequency}
 \acro{HGI}{Horst Görtz Institute for IT Security}
 \acro{HID}{Human Interface Device}
 \acro{HLTA}{HaLT type A}
 \acro{HMAC}{Hash-based Message Authentication Code}
 \acro{HOTP}{HMAC-based One Time Password}
 \acro{HVSP}{High Voltage Serial Programming} 
 \acro{HVG}{High Voltage Generator}
 \acro{HW}{Hamming Weight}
 \acro{I2C}[I$^2$C] {Inter-Integrated Circuit}
 \acro{ICAO}{International Civil Aviation Organization}
 \acro{IC}{Integrated Circuit}
 \acro{ICC}{Integrated Circuit Card}
 \acro{ICSP} {In-Circuit Serial Programming}
 \acro{ID}{Identifier}
 \acro{IDE}{Integrated Development Environment}
 \acro{IFD}{Interface Device}
 \acro{IGBT}{Insulated Gate Bipolar Transistor}
 \acro{ILA}{Integrated Logic Analyzer}
 \acro{ISO}{International Organization for Standardization}
 \acro{ISM}{Industrial, Scientific, and Medical \acroextra{(frequencies)}}
 \acro{IFF}{Identify Friend or Foe}
 \acro{IIR}{Infinite Impulse Response}
 \acro{IP}{Intellectual Property}
 \acro{ISR}{Interrupt Service Routine}
 \acro{ISP}{In-System Programming}
 \acro{IV}{Initialization Vector}
 \acro{IO}{Input-Output}
 \acro{JTAG}{Joint Test Action Group}
 \acro{KDF}{Key Derivation Function}
 \acro{kNN}{$k$-Nearest Neighbors}
 \acro{KPCA}{Kernel Principal Component Analysis}
 \acro{KPA}{Known-Plaintext Attack}
 \acro{KCA}{Known-Ciphertext Attack}
 \acro{LAN}{Local Area Network}
 \acro{LDA}{Linear Discriminant Analysis}
 \acro{LED}{Light-Emitting Diode}
 \acro{LF}{Low Frequency}
 \acro{LFI}{Laser Fault Injection}
 \acro{LFSR}{Linear Feedback Shift Register}
 \acro{LIVA}{Light-Induced Voltage Alteration}
 \acro{LIW}{Listening Window}
 \acro{LSB}{Least Significant Bit}
 \acroplural{LSB}[LSBs]{Least Significant Bits}
 \acro{LSByte}{Least Significant Byte}
 \acro{LUT}{Look-Up Table}
 \acroplural{LUT}[LUTs]{look-up tables}
 \acro{MAC}{Message Authentication Code}
 \acro{MF}{Medium Frequency}
 \acro{MIA}{Mutual Information Analysis}
 \acro{MISR}{Multiple Input Signature Register}
 \acro{MITM}{Man-In-The-Middle}
 \acro{MOSFET}{Metal-Oxide Semiconductor Field-Effect Transistor}
 \acro{MRZ}{Machine Readable Zone}
 \acro{MRTD}{Machine Readable Travel Document}
 \acro{MSB}{Most Significant Bit}
 \acro{MSByte}{Most Significant Byte}
 \acro{nPA}{New German ID Card}
 \acro{NA}{Numerical Aperture}
 \acro{NACK}{Negative Acknowledgment}
 \acro{NDA}{Non-Disclosure Agreement}
 \acro{NIR}{Near Infrared}
 \acro{NIST}{National Institute of Standards and Technology}
 \acro{NLFSR}{Non-Linear Feedback Shift Register}
 \acro{NLF}{Non-Linear Function}
 \acro{NFC}{Near Field Communication}
 \acro{nMOS} {n-channel Metal Oxide Semiconductor}
 \acro{NRZ}{Non-Return-to-Zero \acroextra{(encoding)}}
 \acro{NOP}{No Operation}
 \acro{NVM}{Non-Volatile Memory}
 \acro{OATH}{Initiative of Open Authentication}
 \acro{OBIC}{Optical Beam Induced Current}
 \acro{OBIRCH}{Optical  Beam  Induced  Resistance  Change}
 \acro{ONNX}{Open Neural Network Exchange}
 \acro{OOK}{On-Off-Keying}
 \acro{OP}{Operational Amplifier}
 \acro{OS}{Operating System}
 \acro{OTP}{One-Time Password}
 \acro{PA}{Passive Authentication}
 \acro{PACE}{Password Authenticated Connection Establishment}
 \acro{PC}{Personal Computer}
 \acro{PCA}{Principal Component Analysis}
 \acro{PCB}{Printed Circuit Board}
 \acroplural{PCB}[PCBs]{Printed Circuit Boards}
 \acro{PCD}{Proximity Coupling Device}
 \acro{PDN}{Power Distribution Network}
 \acroplural{PDN}[PDNs]{Power Distribution Networks}
 \acro{PEA}{Photonic Emission Analysis}
 \acro{PICC}{Proximity Integrated Circuit Card}
 \acro{PIT}{Programmable Identification Transponder}
 \acro{PKI}{Public Key Infrastructure}
 \acro{PLL}{Phase Locked Loop}
 \acro{PMS}{Perfectly Masked Squaring}
 \acro{PMM}{Perfectly Masked Multiplication}
 \acro{pMOS} {p-channel Metal Oxide Semiconductor}
 \acro{PPC}{Pulse Pause Coding}
 \acro{PPS}{Protocol and Parameter Selection}
 \acro{PPSR}{Protocol and Parameter Selection Request}
 \acro{PRN}{Pseudo-Random Number}
 \acro{PRNG}{Pseudo-Random Number Generator}
 \acro{PS}{Passive Serial \acroextra{(mode)}}
 \acro{PSK}{Phase Shift Keying}
 \acro{PTT}{Platform Trust Technology}
 \acro{RADAR}{Radio Detection And Ranging}
 \acro{RAM}{Random Access Memory}
 \acro{RATS}{Request for Answer To Select}
 \acro{REQA}{REQuest type A}
 \acro{RISC}{Reduced Instruction Set Computing}
 \acro{RF}{Radio Frequency}
 \acro{RFID}{Radio Frequency IDentification}
 \acro{RGT}{Request Guard Time}
 \acro{RKE}{Remote Keyless Entry}
 \acro{RNG}{Random Number Generator}
 \acro{ROM}{Read Only Memory}
 \acro{RO}{ring oscillator}
 \acroplural{RO}[ROs]{ring oscillators}
 \acro{RSA}{Rivest Shamir and Adleman}
 \acro{RTL}{register transfer level}
 \acro{RTF}{Reader Talks First}
 \acro{RUB}{Ruhr-University Bochum}
 \acro{SAK}{Select AcKnowledge}
 \acro{SAM}{Square-and-Multiply}
 \acro{SCA}{Side-Channel Analysis}
 \acro{SCARE}{Side-Channel Analysis for Reverse-Engineering}
 \acro{SDD}{Small Delay Defect}
 \acroplural{SDD}[SDDs]{Small Delay Defects}
 \acro{SDK}{Software Development Kit}
 \acro{SDR}{Software-Defined Radio}
 \acro{SECT}{Security Transponder}
 \acro{SEI}{Seebeck Effect Imaging}
 \acro{SEM}{Scanning Electron Microscopy}
 \acro{SIKE}{Supersingular Isogeny Key Encapsulation}
 \acro{SMD} {Surface Mount Device}
 \acro{SMF} {Single Mode Fiber}
 \acro{SNR}{Signal to Noise Ratio}
 \acro{SHA}{secure hash algorithm}
 \acro{SHA-1}{Secure Hash Algorithm 1}
 \acro{SLM}{Silicon Lifecycle Management}
 \acro{SMA}{SubMiniature version A \acroextra{(connector)}}
 \acro{SQL}{Structured Query Language}
 \acro{SOF}{Start Of Frame}
 \acro{SOIC}{Small-Outline Integrated Circuit}
 \acro{SPA}{Simple Power Analysis}
 \acro{SPN}{Substitution-Permutation Network}
 \acro{SPOF}{Single Point of Failure}
 \acro{SRAM}{Static Random Access Memory}
 \acro{SVM}{Support Vector Machines}
 \acro{TA}{Template Attack}
 \acro{TAP}{Test Access Point}
 \acroplural{TAP}[TAPs]{Test Access Points}
 \acro{TCB}{Trusted Computing Base}
 \acroplural{TCB}[TCBs]{Trusted Computing Bases}
 \acro{TDC}{Time-to-Digital Converter}
 \acroplural{TDC}[TDCs]{Time-to-Digital Converters}
 \acro{TEM}{Transmission Electron Microscopy}
 \acro{TI}{Threshold Implementation}
 \acro{TIVA}{Thermally-Induced Voltage Alteration}
 \acro{TLU}{Table Look Up}
 \acro{TTF}{Tag Talks First}
 \acro{TMTO}{Time-Memory Tradeoff}
 \acro{TMDTO}{Time-Memory-Data Tradeoff}
 \acro{TMM}{Transformed Multiplicative Masking}
 \acro{TNR}{Trace-to-Noise Ratio}
 \acro{TPM}{Trusted Platform Module}
 \acroplural{TPM}[TPMs]{Trusted Platform Modules}
 \acro{TWI}{Two Wire Interface}
 \acro{SOC}{Start Of Communication}
 \acro{SoC}{System on Chip}
 \acroplural{SoC}[SoCs]{Systems on Chip}
 \acro{SHF}{Superhigh Frequency}
 \acro{SPI}{Serial Peripheral Interace}
 \acroplural{UC}[$\mu$Cs]{Microcontrollers}
 \acro{UART}{Universal Asynchronous Receiver Transmitter}
 \acro{UHF}{Ultra High Frequency}
 \acro{UID}{Unique Identifier} 
 \acro{UMTS}{Universal Mobile Telecommunications System}
 \acro{USB}{Universal Serial Bus}
 \acro{USRP}{Universal Software Radio Peripheral}
 \acro{USRP2}{Universal Software Radio Peripheral (version 2)}
 \acro{UVC} [UV-C] {Ultraviolet-C \acroextra{(light)}}
 \acro{VCC}[$V_{\sf{CC}}$] {Common collector Voltage \acroextra{(general positive supply voltage)}}
 \acro{VDD}[$V_{\sf{DD}}$] {Drain Voltage \acroextra{(at a transistor)}}
 \acro{VDS}[$V_{\sf{DS}}$] {Drain-Source Voltage \acroextra{(at a transistor)}} 
 \acro{VGD}[$V_{\sf{GD}}$] {Gate-Drain Voltage \acroextra{(at a transistor)}}
 \acro{VGS}[$V_{\sf{GS}}$] {Gate-Source Voltage \acroextra{(at a transistor)}}
 \acro{VHDL}{VHSIC (Very High Speed Integrated Circuit) Hardware Description Language}
 \acro{VSS}[$V_{\sf{SS}}$] {Source Voltage \acroextra{(at a transistor)}}
 \acro{WFNA} {White Fuming Nitric Acid}
 \acro{WLAN}{Wireless Local Area Network}
 \acro{WUPA}{Wake-UP A}
 \acro{XOR}{Exclusive OR}
\end{acronym}

\title{MaliGNNoma: GNN-Based Malicious Circuit Classifier for Secure \bluee{Cloud} FPGAs}

\author{
    \IEEEauthorblockN{Lilas Alrahis\IEEEauthorrefmark{1}, Hassan Nassar\IEEEauthorrefmark{2}, Jonas Krautter\IEEEauthorrefmark{2}, Dennis Gnad\IEEEauthorrefmark{2}, Lars Bauer\IEEEauthorrefmark{2}, Jörg Henkel\IEEEauthorrefmark{2} and Mehdi Tahoori\IEEEauthorrefmark{2}}   
    \IEEEauthorblockA{\IEEEauthorrefmark{1}New York University Abu Dhabi, Abu Dhabi, United Arab Emirates (UAE)\\
    \IEEEauthorrefmark{2}Karlsruhe Institute of Technology (KIT), Institute for Computer Engineering (ITEC), Karlsruhe, Germany\\
                      Email: lma387@nyu.edu, \{hassan.nassar, jonas.krautter, dennis.gnad, lars.bauer, henkel, mehdi.tahoori\}@kit.edu}
}

\maketitle
\renewcommand{\headrulewidth}{0.0pt}
\thispagestyle{fancy}
\lhead{}
\rhead{}
\chead{This is the author's version of the work.
The definitive Version of Record will appear in the 2024 IEEE International Symposium on Hardware Oriented Security and Trust (HOST).}
\cfoot{}

\begin{abstract}
\bluee{The security of cloud field-programmable gate arrays (FPGAs) faces challenges from untrusted users attempting fault and side-channel attacks through malicious circuit configurations. Fault injection attacks can result in denial of service, disrupting functionality or leaking secret information. This threat is further amplified in multi-tenancy scenarios. Detecting such threats before loading onto the FPGA is crucial, but existing methods face difficulty identifying sophisticated attacks.}

We present \textit{MaliGNNoma}, a machine learning-based solution that accurately identifies malicious FPGA configurations.  \bluee{Serving as a netlist scanning mechanism, it can be employed by cloud service providers as an initial security layer within a necessary multi-tiered security system.} By leveraging the inherent graph representation of FPGA netlists, MaliGNNoma employs a graph neural network (GNN) to learn distinctive malicious features, surpassing current approaches. To enhance transparency, MaliGNNoma utilizes a parameterized explainer for the GNN, labeling the FPGA configuration and pinpointing the sub-circuit responsible for the malicious classification.

Through extensive experimentation on the ZCU102 board with a Xilinx UltraScale+ FPGA, we validate the effectiveness of MaliGNNoma in detecting malicious configurations, including sophisticated attacks, such as those based on benign modules, like cryptography accelerators. MaliGNNoma achieves a classification accuracy and precision of 98.24\% and 97.88\%, respectively, surpassing state-of-the-art. We compare MaliGNNoma with five state-of-the-art scanning methods, revealing that not all attack vectors detected by MaliGNNoma are recognized by existing solutions, further emphasizing its effectiveness. Additionally, we make MaliGNNoma and its associated dataset publicly available.

\end{abstract}

\section{Introduction}

\bluee{In response to the evolving landscape of high-speed computation demands in the cloud, traditional central processing units (CPUs) and graphics processing units (GPUs) have become inadequate in terms of latency, efficiency, and throughput~\cite{jin2020security}. To address the increasing performance requirements, field programmable gate arrays (FPGAs) have been integrated into cloud computation platforms. This integration allows users to customize their hardware accelerators for computationally intensive tasks in the cloud. For instance, \acp{CSP}, such as Amazon Web Service (AWS)~\cite{AWSF1} and Microsoft~\cite{azure}, offer cloud-based FPGAs that clients can rent and customize with their own logic~\cite{fpgacloud}; this model can be referred to as \textit{FPGA-as-a-Service} (FaaS).}

\bluee{As clients have control over programming these FPGAs, multiple security concerns are raised. One prominent example is \textit{fault-injection attacks}, wherein circuits are intentionally configured to induce severe voltage fluctuations, leading to hardware faults.} \rede{Such faults can either be subtle and cause computation errors, or cause a crash of the FPGA device and therefore \ac{DoS}~\cite{Gnad2017voltage}, jeopardizing the entire availability of the system.} \rede{When evaluating the financial loss for commercial \acp{CSP} versus the costs of performing the attack, the cost of the downtime of FPGA infrastructure is 10$\times$ as much as the costs for the attacker causing the downtime~\cite{La2021}. Such powerful attacks put the \acp{CSP} at a severe disadvantage to the attackers and make them potential victims of sabotage~\cite{La2021}.}

\bluee{These threats can be even more amplified with the concept of \textit{multi-tenancy.} To meet the performance demands of the FaaS model, large and expensive FPGAs are deployed in the cloud. To enhance cost-efficiency and maximize return on investment for CSPs, it is essential to fully utilize this capacity.
When clients with the necessary resource and computing demands are not present, the FPGA fabric can be shared among multiple clients (\textit{tenants}) with smaller resource requirements.}
\rede{Sharing the fabric optimizes the FPGA utilization and makes it more cost-effective for individual users, if less users are capable of utilizing the entire FPGA fabric~\cite{vaishnav2018survey}.}

 \begin{figure}[!t]
\centering
\includegraphics[width=0.49\textwidth]{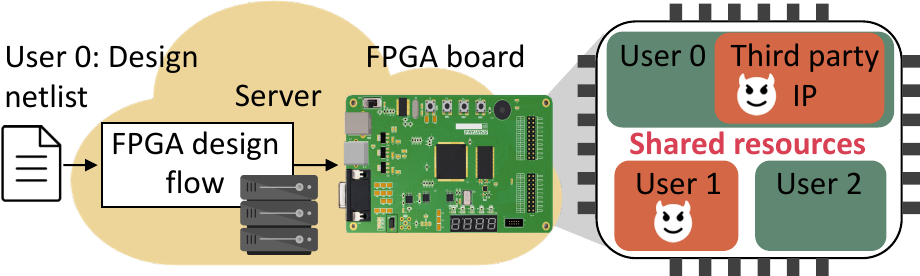}
\caption{\bluee{Cloud} FPGA with shared resources and open attack vectors. The threat model involves \bluee{either a malicious user utilizing the entire FPGA or a malicious third-party IP provider, or in the context of multi-tenancy, a malicious co-tenant.} Assuming the cloud provider is trustworthy, malicious entities can manipulate their designs intentionally, leading to fault-injection attacks.}
\label{fig:intro}
\end{figure}
\bluee{Multi-tenancy has not been realized commercially yet, mainly due to security challenges.
One of the reasons is the shared power delivery networks of the FPGA fabric among tenants, which can create potential attack vectors~\cite{Krautter2019mitigating}.}
\rede{Next to the fault-injection attacks that can crash the FPGAs, \textit{side-channel attacks} use indirect power measurements to exfiltrate secret information from one tenant's FPGA fabric through a malicious co-tenant~\cite{schellenberg2018inside,ramesh2018fpga,Moini2021BNN}.
Additionally, more fine-grained fault-injection attacks do not only affect the availability of resources, but can also affect data integrity by causing delicate faults that only affect individual bytes. Using that, attacks exist to extract cryptographic secrets~\cite{remotefault, Krautter2018} or information in neural network accelerators~\cite{boutros20}.}
\bluee{Further research on this topic is needed to reveal potentially more vulnerabilities and provide effective countermeasures, making multi-tenancy secure and hence feasible. Fig.~\ref{fig:intro} demonstrates the scenario of cloud FPGAs that support multi-tenancy, with the accompanying threat model explanation.}

\rede{\textbf{In this work}, we address the security of cloud FPGAs against fault-injection attacks, while considering multi-tenancy. Side-channel attacks are not addressed here, but can be solved orthogonal to us -- for instance on the level of timing analysis, as suggested in~\cite{Krautter2019mitigating}.}
\bluee{While we consider multi-tenancy scenarios, our research is not entirely confined to that aspect.}

In this context, researchers have explored various protection schemes to enhance the security of cloud FPGA deployments, including architectural modifications to the FPGAs and circuit design enhancements to mitigate vulnerability to fault-injection and side-channel attacks~\cite{activefences, ahmadi2023fpga}. Nevertheless, ensuring the security of cloud FPGAs necessitates a multi-tiered defense system that integrates all the aforementioned measures for maximum security. Additionally, the first layer of security in such a system should proactively detect and prevent attacks before loading the user's configuration to the FPGA, thus preventing any potential damage. One \blue{effective} approach involves a \textit{hypervisor} that checks and allows only benign bitstreams to be loaded onto the FPGA fabric~\cite{zeitouni21} (we refer to such methods as scanning techniques). However, existing scanning methods have serious limitations, as discussed next.

\subsection{State-of-the-Art and their Limitations}

\textbf{Scanning the Bitstream:}
Existing methods either reverse-engineer the netlist from the bitstream~\cite{Krautter2019mitigating,la2020fpgadefender}, or scan the bitstream directly for malicious signatures~\cite{chaudhuri2023diagnosis}. However, in the workflow of cloud FPGAs, users provide a design prior to the bitstream, i.e., a Verilog netlist, out of which the final bitstream needs to be generated by the cloud-hosted electronic design automation (EDA) tools, which also perform respective \ac{DRC}~\cite{La2021}. Hence, incorporating a scanning method at \ac{DRC} level directly fits into the working model of \acp{CSP}—that is, the methods should be tailored to scan the netlist itself directly instead of scanning the bitstream. Please note that a bitstream can only be seen as a more obscured netlist~\cite{bitman}, it does not provide proper security against intellectual property (IP) theft. A properly encrypted bitstream is indistinguishable from random data, and thus, cannot be checked for security violations.

\textbf{Limited Security:} verifies the netlist against a set of predefined rules to detect any potential security vulnerabilities or design errors.
However, the existing approaches adopted by \acp{CSP} like Amazon, particularly those based on Xilinx FPGAs, primarily focus on detecting various types of \acp{RO}. As a result, they fail to address more sophisticated and complex \textit{attack vectors},\footnote{These attacks employ carefully combined cryptographic cores and power-intensive modules for fault injection.} limiting their effectiveness in ensuring comprehensive security for multi-tenant FPGAs. Furthermore, existing scanning techniques, such as \textit{FPGAdefender}~\cite{la2020fpgadefender} or the work in~\cite{Krautter2019mitigating}, are only designed for a limited set of components or very simple FPGAs, which means only elements such as \acp{LUT} and basic arithmetic blocks are considered, but not sequential-based attacks~\cite{Provelengios2020}.
The capabilities of existing scanning methods in detecting the different types of fault-injection attacks are summarized in Table~\ref{tab:comparison}. These attacks will be explained in detail in Section~\ref{sec:background}. In this work, we extend the scope of applicability to all available FPGA resources.

{\textbf{Identifying Malicious Modules:} Labeling the netlist/bitstream as malicious may lack insights and justification. Methods that classify the netlist/bitstream while providing pointers to specific malicious parts are valuable for design analysis. Table~\ref{tab:comparison} (right-most column) illustrates how not all existing techniques offer this feature.}

{\textbf{Machine Learning (ML)-based Detection:}} \blue{Recently, ML techniques have shown promise in scanning bitstreams to identify malicious configurations~\cite{CNNbased}. However, there is a research gap in investigating netlist-level approaches for FPGA security, as current ML research focuses on scanning the bitstream directly, proving insufficient for use with the CSP workflow. Moreover, it does not provide justification for identifying malicious parts within the configuration. This lack of transparency poses a challenge in offering feedback for the next layer of scrutiny and verification.}

\begin{table}[!t]
\centering
\caption{\textsc{Capabilities Offered by FPGA Scanning Techniques.}}
\label{tab:comparison}
\resizebox{0.49\textwidth}{!}{%
\begin{tabular}{cccccc}
\hline
\textbf{Techniques} & \textbf{\begin{tabular}[c]{@{}c@{}}Self-oscil.\\ attacks\end{tabular}} & \textbf{\begin{tabular}[c]{@{}c@{}}Hidden\\ attacks\end{tabular}} & \textbf{\begin{tabular}[c]{@{}c@{}}Sequen.\\ attacks\end{tabular}} & \textbf{\begin{tabular}[c]{@{}c@{}}Partial\\ designs\end{tabular}} & \textbf{\begin{tabular}[c]{@{}c@{}}Labeling\\ sub-circuits\end{tabular}} \\\hline
Tool of~\cite{learnmal23} & \color{cadmiumgreen}{\cmark} & \color{cadmiumgreen}{\cmark} & \color{red}{\xmark} & \color{red}{\xmark} & \color{red}{\xmark}
\\ \hline
Tool of~\cite{CNNbased} & \color{cadmiumgreen}{\cmark} & \color{red}{\xmark} & \color{red}{\xmark} & \color{cadmiumgreen}{\cmark} & \color{red}{\xmark}
\\ \hline
Tool of~\cite{chaudhuri2023diagnosis} & \color{cadmiumgreen}{\cmark} & \color{red}{\xmark} & \color{cadmiumgreen}{\cmark} & \color{red}{\xmark} & \color{red}{\xmark}
\\ \hline
Tool of~\cite{la2020fpgadefender} & 
\color{cadmiumgreen}{\cmark} & \color{red}{\xmark} & \color{red}{\xmark} & \color{cadmiumgreen}{\cmark} & \color{cadmiumgreen}{\cmark}
\\ \hline
Tool of~\cite{Krautter2019mitigating} & 
\color{cadmiumgreen}{\cmark} & \color{red}{\xmark} & \color{red}{\xmark} & \color{cadmiumgreen}{\cmark} & \color{red}{\xmark}
\\ \hline
\textbf{MaliGNNoma} & 
\textbf{\color{cadmiumgreen}{\cmark}} & 
\color{cadmiumgreen}{\cmark} & 
\color{cadmiumgreen}{\cmark} & 
\color{cadmiumgreen}{\cmark} & 
\color{cadmiumgreen}{\cmark} 
\\ \hline
\multicolumn{6}{c}{{\color{cadmiumgreen}{\cmark}} denotes detection/labeling capability, while {\color{red}{\xmark}} indicates incapability}\\
\end{tabular}
}
\end{table}
\subsection{Research Challenges Addressed in this Work}
\blue{ML's capacity to learn attack signatures and generalize to unseen designs is a significant advantage that can lead to addressing the limitations of state-of-the-art scanning solutions. However, it has not been developed to its fullest potential in this context. Our goal is to leverage ML for FPGA netlist scanning, a promising approach that presents a unique set of challenges, which we discuss next.}
\begin{enumerate}
 \item Traditional ML models are designed to handle fixed-size numerical vectors/matrices. However, netlists inherently possess a non-Euclidean representation and cannot be mapped to vectors without performing feature engineering. Manual feature engineering is prone to errors and highly dependent on the specific dataset at hand. Furthermore, there is no guarantee that it will result in an optimized representation of the data. Hence, it is necessary to develop ML models that can automatically extract optimized vector representations from complicated netlist structures. These representations can then be used for the learning process. Recently, graph neural networks (GNNs) have emerged as a new type of ML that operates on graph-structured data. These networks can automatically learn and represent graphs as \textit{vector embeddings}, capturing each graph's important properties in a trainable manner. Since circuits can naturally be represented as graphs, GNNs have achieved superior performance in solving various circuit-related tasks, including hardware security (e.g., detecting IP piracy~\cite{gnn4ip} and functional reverse engineering~\cite{gnnre,appgnn}) and reliability (e.g., predicting circuit delay degradation due to aging~\cite{GNN4IC}). However, GNNs have not been employed yet in the context of FPGA security.

 \item Using ML models as black-box solutions for critical tasks raises \textit{trust and transparency challenges}. Without insights into how and why decisions are made, trusting the model's results can be a concern, impacting the reliability of the classification. 
\end{enumerate}

\subsection{Our Concept and Contributions}
\begin{figure}[!t]
\centering
\includegraphics[width=0.49\textwidth]{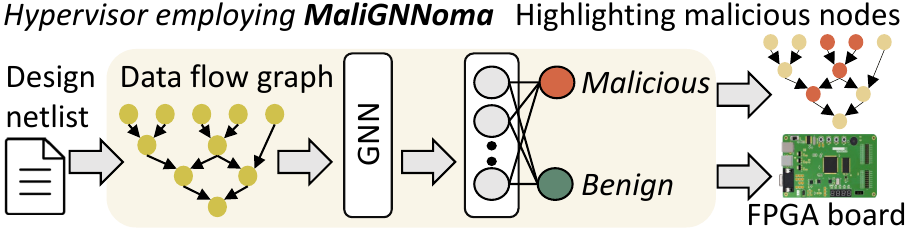}

\caption{High-level view of our work: Identifying malicious FPGA configurations at the netlist-level using graph neural networks (GNNs).}
\label{fig:idea}
\end{figure}

To address the above challenges, we propose the \textit{\textbf{MaliGNNoma framework}} that identifies different netlist-level malicious FPGA configurations. MaliGNNoma leverages GNNs to accurately classify netlists by capitalizing on their inherent graph structure, as depicted in Fig.~\ref{fig:idea}. Unlike existing methods that focus on checking bitstreams, MaliGNNoma directly operates on netlist graphs. This enables the model to preserve semantic information, extract crucial features, and improve detection capabilities. Additionally, MaliGNNoma aligns seamlessly with the workflow of \ac{CSP}, where netlists serve as the primary input for generating bitstreams. MaliGNNoma employs the following techniques:
\begin{enumerate}
\item We present an \textbf{ML-based approach for FPGA netlist checking} that employs a GNN. In our evaluation, we consider two different GNN architectures and investigate the incorporation of attention layers and dropout. Our aim is to provide insights into the capabilities of GNNs for this task, as this is their first application in such a context. We have constructed our own dataset of FPGA netlists to train the GNN, which we to the research community at the following repository: \href{https://github.com/hassanassar/MaliGNNoma}{https://github.com/hassanassar/MaliGNNoma}.

\item \textbf{Interpretation:} Additionally, we integrate our framework with a state-of-the-art parameterized explainer for GNNs~\cite{luo2020parameterized}. This explainer, when given a graph (i.e., a netlist), extracts a subgraph (i.e., sub-circuit) that has the most influence on the GNN's predictions. By identifying specific gates contributing to the prediction of a `malicious' label, MaliGNNoma offers valuable insights for investigation by circuit designers.

\end{enumerate}

\textbf{Key Results:} The effectiveness of MaliGNNoma is showcased by training it over a dataset comprising of benign and malicious designs. The benign benchmarks are collected from ISCAS benchmark~\cite{iscas89}, Berkley benchmark~\cite{2007benchmarking}, Groundhog benchmark~\cite{Groundhog10}, and OpenCores~\cite{OpenCores}. 
The malicious designs contain state-of-the-art attacks as described in~\cite{la2020fpgadefender,learnmal23,Provelengios2020,RamJam19}. We crafted the malicious designs based on state-of-the-art methodologies, as they are not open-sourced. We evaluated all designs in our dataset on a ZCU102 FPGA Board and confirmed that malicious designs injected faults into neighboring tenants.\footnote{The widley-adopted Trust-HUB dataset~\cite{px6s-sm21-22} is for detecting hardware Trojans in application-specific integrated circuits (ASIC). However, we focus on identifying fault-injection attacks in FPGAS (different scope). Integrating this dataset into MaliGNNoma would not be aligned with our objectives.} MaliGNNoma achieves a classification accuracy and precision of 98.24\% and 97.88\%, respectively. \textbf{We make MaliGNNoma and associated benchmarks (benign and malicious) available online.}\footnote{
Code:\href{https://github.com/DfX-NYUAD/MaliGNNoma}{https://github.com/DfX-NYUAD/MaliGNNoma}\\
Dataset:\href{https://github.com/hassanassar/MaliGNNoma}{https://github.com/hassanassar/MaliGNNoma}.}

\begin{figure*}[!t]
\centering
\includegraphics[width=0.95\textwidth]{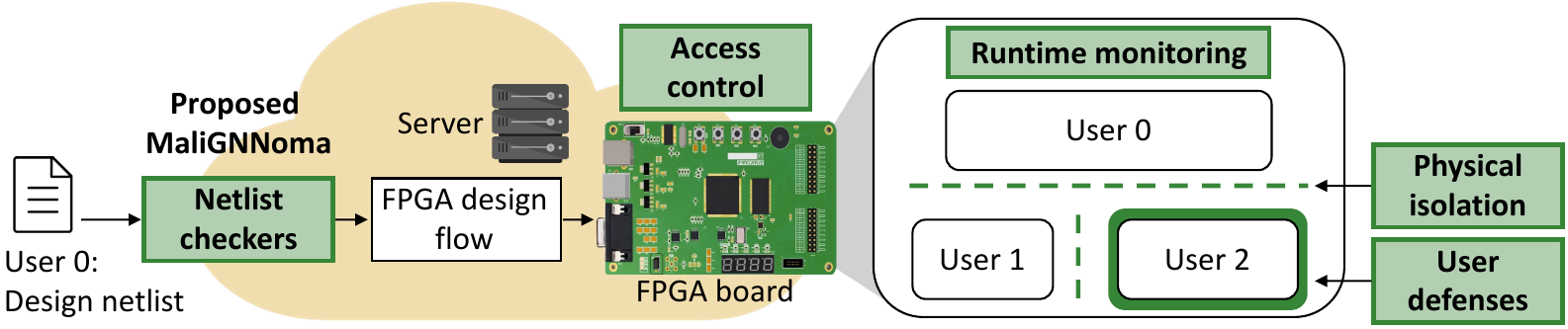}

\caption{Multi-tiered defense system for securing cloud FPGAs. Defense mechanisms, highlighted in green, include netlist checking, access control, physical isolation, and runtime monitoring implemented by the CSP. Users can further enhance security with design-level defenses such as masking and active measures.}
\label{fig:MaliGNNoma_background}
\end{figure*}
\section{Background and Related Work}
\label{sec:background}
\begin{table}[!t]
\centering
\caption{\textsc{Commonly Used Abbreviations and Notations}}
\label{tab:terns}
\resizebox{0.49\textwidth}{!}{%
\begin{tabular}{cc|cc}
\toprule
Term & Definition & Term & Definition \\ \midrule 
 ${h}_v^{(l)}$ & Node $v$ embedding at GNN layer $l$ & ${N}(v)$ & Neighbors of node $v$, excluding $v$ \\ 
 $\sigma$& Nonlinear activation function & $\mathbf{H}^{(l)}$& Node embeddings matrix\\ 

 ${a}_v^{(l)}$ & Aggregated information &$L$ & Total number of GNN layers \\ 
$\mathbf{X}$& Node features matrix & $\mathbf{W}$ & Trainable weight matrix \\
$\mathbf{A}$ & Adjacency matrix &MLP & Multi layer perceptron\\ 
$\mathbf{\widehat{D}}$ & Diagonal degree matrix & $G$ & Graph \\
$h_G$ & Graph-level embedding & E& Edges in $G$\\
$V$ & Nodes in $G$ & $N$ & Number of nodes in $G$\\
$\hat{y}_G$ & Prediction for $G$ & $y_G$ & Ground truth for $G$\\
\bottomrule

\end{tabular}%
}
\end{table}

Ensuring the security of cloud FPGAs requires a multi-tiered defense system, as demonstrated in Fig.~\ref{fig:MaliGNNoma_background}. This system includes netlist checking for malicious implementations, access control between the FPGA processor and peripherals, physical isolation between tenants, and runtime monitoring to detect any suspicious activities~\cite{jin2020security}. All of these defenses should be supported and implemented by the CSP.
Additionally, users can incorporate design-level defenses, such as masking and active defenses, to protect their own designs~\cite{activefences}. All these measures are required to ensure the security of the system against various attacks. Our primary focus is on netlist checking, the first stage of protection.

In this section, we present the necessary background on FPGA-internal attacks, existing countermeasures against them, and information on GNNs, netlist to graph conversion, and the utilization of GNNs for various circuit-related tasks. Table~\ref{tab:terns} provides the notations used throughout the paper.

\subsection{FPGA-Internal Attacks}

Most known FPGA-internal attacks are based on voltage fluctuations or cross-coupling between logically isolated designs, which can be exploited through fault-injection or side-channel attacks, that traditionally have been carried out with test \blue{and} measurement equipment, \blue{while having} physical access.
On one hand, voltage sensors based on FPGA logic can be used to estimate voltage fluctuations, which are then analyzed to recover secret key information~\cite{Zhao2018SCA,schellenberg2018inside} or more recently also information about neural network accelerators~\cite{Moini2021BNN,zhang2021stealing, boutros20}. 
On the other hand, fault-injection attacks cause errors in computation, which can lead to \ac{DoS} as well as \ac{DFA} to extract confidential data~\cite{Gnad2017voltage,Krautter2018}. We focus on fault-injection attacks, which cause a very high voltage drop through excessive switching activity, that is sufficient to cause timing violations in designs integrated in other partitions of the FPGA.

The first FPGA-internal fault-injection attacks were typically implemented using a large number of \acp{RO} (self-oscillating attacks from Table~\ref{tab:comparison}) that need a combinational loop~\cite{Gnad2017voltage,Krautter2018}.
Since cloud platforms such as Amazon EC2 F1 integrate a basic check for combinational loops, stealthy RO variants were developed, making detection difficult in the toolchain~\cite{sugawara2019oscillator}.
Later it was shown that faults can also be caused using access conflicts based on block memory (BRAM) in Xilinx FPGAs~\cite{RamJam19}, and specific ways to configure logic go under the name of glitch amplification~\cite{Glitch20} to also cause very high voltage variations.
In the end, even seemingly benign circuits (sequential attacks from Table~\ref{tab:comparison}) have been shown able \blue{of causing} a sufficient voltage drop \blue{that results} in faulty behavior, which is very stealthy and harder to detect~\cite{Provelengios2020,krautter2021remote}.
To increase the stealthiness of the attack, the attackers mix benign circuits with malicious circuits within the same design (hidden attacks from Table~\ref{tab:comparison})~\cite{learnmal23} which makes it \blue{also} harder to detect.

\subsection{Existing Countermeasures Against FPGA-Internal Attacks}

To prevent FPGA-internal attacks, traditional countermeasures against fault-injection and side-channel attacks can be employed for cryptographic implementations, but come with respectively high overheads.
Instead, some runtime countermeasures are more tailored to the needs of FPGAs, such as \textit{hiding} against side-channel attacks~\cite{activefences} or actively trying to prevent fault-injection attacks~\cite{loopbreaker}, with mixed effectivity. 

Thus, more research was dedicated towards detecting malicious designs before they even get loaded to the FPGA, similar to an \textit{anti-virus} software.
Indeed, these existing works also utilize mechanisms similar to detecting malware in software, but \blue{instead of code patterns search for} patterns in a bitstream \blue{that are typical for either} side-channel or fault-injection attacks~\cite{Krautter2019mitigating}.
However, the difficulty is to detect attacks that are based on seemingly benign circuits, for which~\cite{Krautter2019mitigating} only considered high-fanout as a potential red flag.
Later, \cite{la2020fpgadefender} focused on fault-injection attacks and the various \blue{types} of malicious power-wasting circuits, but without the goal of detecting variants based on benign circuits.

What these two works have in common~\cite{Krautter2019mitigating,la2020fpgadefender}, is that they are based on reverse engineering and checking the bitstream.
Instead, the work in \cite{chaudhuri2023diagnosis} tries to be independent of that, and applies metrics to directly analyze a bitstream for fault-injection attacks.
However, it is debatable if an analysis \blue{on the raw} bitstream {is useful}, since either cloud providers would be able to get sufficient information from FPGA vendors, or the toolchain itself will have to check it before bitstream generation, which is already enforced by the tool flow needed for the Amazon platform~\cite{La2021}.

\subsection{Graph Neural Networks (GNNs)}

A GNN operates by taking a given graph $G$ and encoding it into a vector representation specifically tailored to suit the intended task. In this process, each node within the graph gets represented as a vector known as an ``embedding.'' These embeddings are computed in such a way that nodes with similarities in the graph should ideally be placed close to one another in the embedding space, indicating proximity in terms of distance. The feature vectors assigned to all nodes serve as their initial embedding vectors. Subsequently, the GNN \blue{performs} ``neighborhood aggregation,'' where each node collects and incorporates messages (i.e., embeddings) from its direct neighbors. These incoming messages are then used to update the node's own embedding based on the received information~\cite{kipf2016semi}. These embeddings are crucial for downstream tasks, such as graph classification or other applications, where the learned representations can help in making informed decisions based on the graph structure and its features.

Since circuit netlists resemble graph structures, there has been a growing interest in employing GNNs to solve various circuit problems related to hardware security~\cite{GNN4Sec,omla,appgnn,gnnre,gnn4ip,GNN4TJ,muxlink,untangle,alrahis2023split,mankali2022titan}, reliability~\cite{GNN4REL}, and EDA~\cite{survey_gnn}, primarily focusing on ASIC and analog circuits~\cite{GNN4IC}.

In the following sections, we discuss the process of converting a Verilog netlist into a graph (Section~\ref{sec:back_netlist2graph}), provide a detailed explanation of the neighbor aggregation operation used in GNNs (Section~\ref{sec:back_gnn}), \blue{and show how GNN results can be explainable to a human understanding (Section~\ref{sec:gnn_explanation}).}

\subsubsection{Netlist to Graph Conversion}
\label{sec:back_netlist2graph}
There are several open-source tools that can read a Verilog netlist and convert it into a graph. Specifically, \textit{PyVerilog} is an open-source hardware design processing toolkit that parses Verilog and can construct various graphs, such as data flow graphs (DFGs) and abstract syntax trees (ASTs). This tool is also utilized in state-of-the-art methods for detecting ASIC hardware Trojans using GNNs~\cite{hw2vec,GNN4TJ}.

The process begins with converting the netlist into an AST using the yet another compiler-compiler (yacc) lexical analyzer. An AST is a tree-based graph with nodes representing entities such as operators (mathematical, gates, loop, conditional), signals, or signal attributes. The edges represent the relationships between these nodes. To generate the DFG, a module analyzer first extracts a list of modules from the AST. Each module definition includes a signal list for inputs, outputs, and parameters. 

Next, the signal analyzer traverses the AST again to collect signal definitions. Signal declarations are analyzed, and the scope of each signal is recorded. Simultaneously, an optimizer resolves constant value definitions, such as parameters and local parameters. Module parameters can be overwritten by their parent module, and the actual hardware structure dependent on these parameters becomes fixed once they are determined.

By utilizing the analysis results from the optimizer, the actual parameters for each module instance and the entire module hierarchy are determined. Finally a bind analyzer generates a DFG for each signal, with each signal serving as the root node. Finally, the signal DFGs are merged to form the resulting graph denoted as $G=(V, E)$. The DFG represents data dependencies, where nodes in $V$ correspond to signals, constant values, and operations like XOR, AND, and concatenation. The edges $E$ depict the data dependency relations between pairs of nodes. The adjacency matrix $\mathbf{A}\in \{0,1\}^{N\times N}$ is constructed from $E$, where $\mathbf{A}_{u,v}=1$ iff $(i,j)\in E$. $N$ denotes the number of nodes in the graph, i.e., $|V|$.

\subsubsection{GNN Architectures}
\label{sec:back_gnn}

The GNN neighborhood aggregation operation involves two main functions: \texttt{Aggregate} and \texttt{Update}, which are repeated for a pre-determined number of layers $L$. During each iteration $l$, the process updates the node embeddings matrix $\mathbf{H}^{(l)}$ based on the node representations $h_v^{(l-1)}$ from $\mathbf{H}^{(l-1)}$, where $\mathbf{H}^{(0)}$ is the same as the feature matrix $\mathbf{X}$, as given by:

\begin{equation}
 a_v^{(l)} = \texttt{Aggregate}^{(l)}(\{h_u^{(l-1)}: u \in N(v)\})
\end{equation}
\begin{equation}
 h_v^{(l)} = \texttt{Update}^{(l)}(h_v^{(l-1)}, a_v^{(l)} )
\end{equation}

For each node $v$, the embedding after $l$ iterations is denoted as $h_v^{(l)} \in \mathbb{R}^{C^l}$, where $C^l$ represents the embedding dimension at layer $l$. The \texttt{Aggregate} operation collects the features, i.e., messages, of the neighboring nodes $N(v)$ to generate an aggregated feature vector $a_v^{(l)}$ for layer $l$. The \texttt{Update} operation then combines the previous node embedding $h_v^{(l-1)}$ with $a_v^{(l)}$ to produce an updated embedding vector $h_v^{(l)}$. This updated embedding captures the properties of the node $v$ itself and also the properties of its neighborhood $N(v)$. The final propagated node embeddings after L iterations are denoted using the matrix $\mathbf{H}^{L}$. GNN implementations differ based on the employed \texttt{Aggregate} and \texttt{Update} functions. In our work, we consider the standard graph convolutional neural network (GCN)~\cite{kipf2016semi} and the graph isomorphism network (GIN)~\cite{xu2018powerful}. Their implementations are explained next, starting with the GCN.

\textbf{GCN:} In each iteration $l$, 
the embedding matrix $\mathbf{H}^{(l)}$ will be updated as follows;
\begin{equation}
 \mathbf{H}^{(l)} = \sigma(\mathbf{\widehat{D}}^{-\frac{1}{2}} \mathbf{\widehat{A}} \mathbf{\widehat{D}}^{-\frac{1}{2}} \mathbf{H}^{(l-1)} \mathbf{W}^{(l-1)})
\end{equation}

The matrix $\mathbf{\widehat{A}}$ is obtained by adding the identity matrix $\mathbf{I}$ to the original adjacency matrix $\mathbf{A}$. This addition of self-loops enables the incorporation of the previously computed embeddings of the target nodes. The matrix $\mathbf{\widehat{D}}$ is a diagonal degree matrix used for normalizing $\mathbf{\widehat{A}}$. The activation function $\sigma(.)$ represents a non-linear activation function, such as the rectified linear unit (ReLU). The matrix $\mathbf{W}^{(l-1)}$ is a trainable weight matrix. 

\begin{figure*}[!t]
\centering
\includegraphics[width=0.98\textwidth]{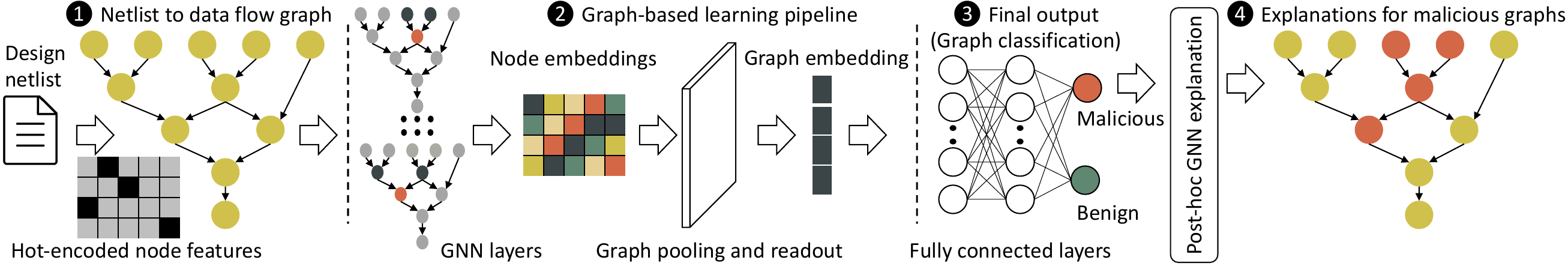}

\caption{Proposed MaliGNNoma methodology: Synthesized netlists are transformed into DFGs for training a GNN to separate malicious and benign samples. A GNN explanation tool identifies nodes contributing to malicious classification.}
\label{fig:MaliGNNoma_flow}
\end{figure*}

\textbf{Graph Classification:} In order to perform graph classification, an overall graph-level embedding must be extracted by applying an order invariant \texttt{READOUT} function, such as \texttt{Sum}, \texttt{Max}, or \texttt{Average}, on the node embeddings $\mathbf{H}^{(L)}$. We denote the graph embedding for each graph $G$ as $h_G$, which is used to make a prediction $\hat{y}_G$ using a multilayer-perceptron (MLP) stage about the graph. For example, the GNN can be trained to minimize the cross-entropy loss for training graphs, as follows. Where $y_G$ denotes the ground truth label.
\begin{equation}
Loss(\{y_G\}, \{\hat{y}_G\}) = \sum_{G} y_G * log_e(\hat{y_G}),\label{loss:cross}
\end{equation}

A common practice involves the addition of a pooling layer preceding the graph \texttt{READOUT} stage. This step focuses on selecting a subset of nodes for graph representation, as opposed to the simultaneous processing of all nodes. Specifically, in an \texttt{Attention}-based pooling layer, the top-$k$ filtering technique can be applied based on the nodes' \texttt{Attention} scores. This procedure entails identifying and retaining the top-k nodes with the highest scores while discarding the rest.

To obtain the scores for the filtering process, one approach is to employ a separate trainable GNN layer that takes into account both the node features and the topological characteristics, as done in~\cite{hw2vec,gnnre}.

\textbf{GIN:} The GIN updates node representations as follows.

\begin{equation} 
h_v^{(l)} = {\texttt{MLP}}^{(l)} \left( h_v^{(l-1)} + \sum\nolimits_{u \in N(v)}{h_u^{(l-1)}} \right)
\end{equation}

GIN considers all structural information from all iterations of the model by concatenating graph embeddings across all layers of GIN as follows. The \texttt{READOUT} function adds all node embeddings from the same layer. 
\begin{equation}
h_{G} = \texttt{CONCAT} \left( \texttt{READOUT} \left( \left\lbrace h_v^{(l)} | v \in G \right\rbrace \right) \ \big\vert \ l = 0,1, \ldots, L \right)
\end{equation}

\subsubsection{GNN Explanation}\label{sec:gnn_explanation}
As with most neural networks, the process through which GNNs reach their classifications lacks a straightforward explanation for human understanding. Addressing this challenge, the authors in~\cite{ying2019gnnexplainer} introduced \textit{GNNExplainer} that implements an optimization task aimed at maximizing the mutual information between a GNN's prediction and a distribution of potential subgraph structures. Hence, GNNExplainer can identify the specific subgraph and node structures contributing to a particular classification.
Expanding upon the concepts introduced in~\cite{ying2019gnnexplainer}, D. Luo \textit{et al.} have extended GNNExplainer and introduced \textit{PGExplainer}~\cite{luo2020parameterized}. PGExplainer is a versatile parameterized explainer designed to be applicable to a broad spectrum of GNN-based models. 
PGExplainer adopts a deep neural network to parameterize the generation process of explanations. Compared to existing work, PGExplainer exhibits superior generalization capabilities. Similar to GNNExplainer, PGExplainer provides a subgraph that influenced the GNN's prediction.

The quality of an explanation can be assessed using various metrics. One of the most common evaluation metric is fidelity- (detailed in~\cite{amara2022graphframex}). Fidelity- evaluates the contribution of the generated explanatory subgraph to the initial prediction, by presenting only the subgraph to the model.

\section{Proposed MaliGNNoma Methodology}
\label{sec:MaliGNNoma}

In this section, we discuss our proposed MaliGNNoma framework in detail, as illustrated in Fig.~\ref{fig:MaliGNNoma_flow}. MaliGNNoma takes a post-synthesis netlist as input and converts it into a graph representation, effectively transforming the problem of identifying malicious circuits into a graph classification task. Additionally, it provides a list of influential nodes in the graph contributing to the prediction of a \textit{``malicious''} label.
Next, we discuss the netlist to graph conversion and the GNN learning steps. Subsequently, we will present the dataset generation process, which significantly impacts the success of the task.
\subsection{MaliGNNoma Pipeline}

\subsubsection{Netlist to DFG} MaliGNNoma employs the open-source PyVerilog tool to parse the Verilog netlists and convert them into DFGs (see Fig.~\ref{fig:MaliGNNoma_flow}~\Circled{\scriptsize\textbf{1}}), as explained in Section~\ref{sec:back_netlist2graph}. MaliGNNoma assigns a one-hot encoded feature to each node in the DFG to indicate its type. The feature vector length per node is $37$, representing all possible types of nodes in a DFG. $\mathbf{X}$ represents the nodes feature matrix.

\subsubsection{GNN-based Learning} The graph $G$ is passed through the employed graph-based learning pipeline, which consists of graph convolution layers, graph pooling layers, and graph \texttt{READOUT} (see Fig.~\ref{fig:MaliGNNoma_flow}~\Circled{\scriptsize\textbf{2}}). In our evaluation, we consider both the GCN and the GIN architectures explained in Section~\ref{sec:back_gnn}.

\begin{figure*}[!t]
\centering
\includegraphics[width=0.96\textwidth]{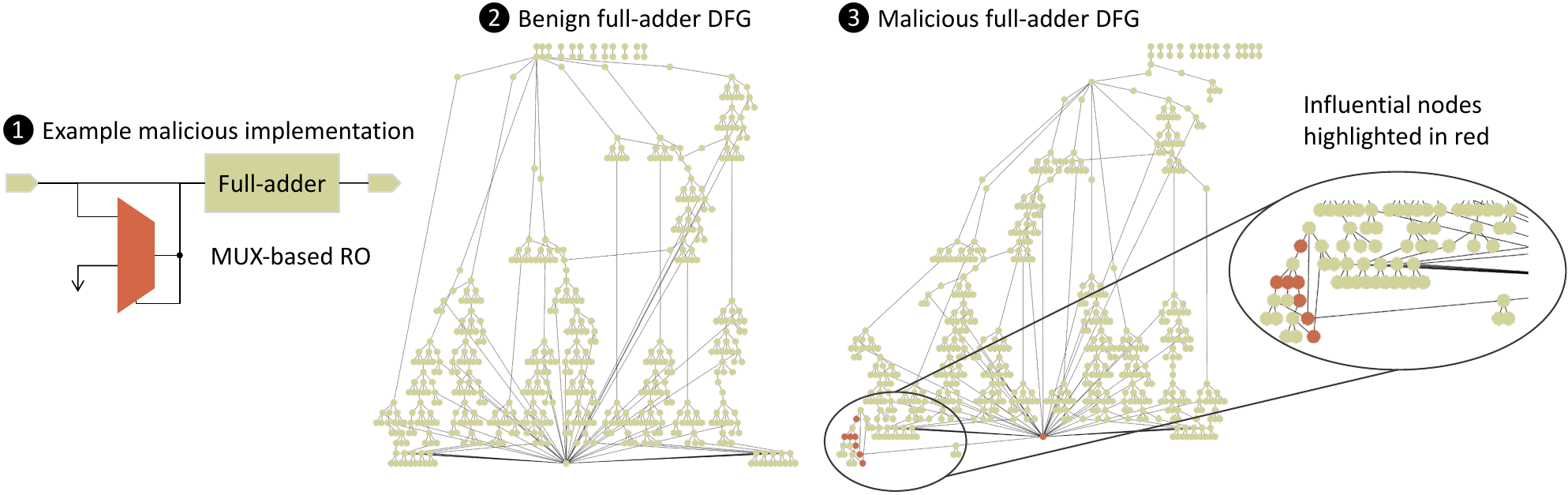}
\caption{\Circled{\scriptsize\textbf{1}}~Example of malicious full-adder with a MUX-based RO connected to its input. \Circled{\scriptsize\textbf{2}} The DFG of a benign full-adder without the MUX-based RO. \Circled{\scriptsize\textbf{3}} The DFG with the MUX-based RO included. \Circled{\scriptsize\textbf{3}} Also illustrates the output of the GNN explainer, highlighting the malicious nodes in the DFG, which correspond to the MUX-based RO.}
\label{fig:dfg_example}
\end{figure*}

\subsubsection{Localizing the Malicious Nodes} By utilizing the \textit{PGExplainer}, our model can identify the influential nodes contributing to the prediction of a \textit{``malicious''} label. 

\blue{Fig.~\ref{fig:dfg_example}~\Circled{\scriptsize\textbf{1}} showcases a high-level implementation of a full-adder with a MUX-based RO connection. The benign full-adder's DFG is depicted in~\Circled{\scriptsize\textbf{2}} and classified as benign by MaliGNNoma. In contrast,~\Circled{\scriptsize\textbf{3}} represents the DFG of the malicious full-adder with the MUX-based RO connection, classified as malicious by MaliGNNoma. Notably, MaliGNNoma goes beyond classification and highlights the most influential nodes (colored in red), corresponding to the MUX-based RO. This automated process aids in analyzing netlists and identifying malicious components, even in cases involving third-party IPs incorporated into the design.} \bluee{Please note that if a design has only one RO (such as this adder example), it will not lead to fault injection. However, MaliGNNoma will flag it, following the industry standard set by Xilinx and AWS tools. We aim to maintain a similarly strict policy. Yet, MaliGNNoma can be retrained to take into account the number of ROs when making the final decision, if required.
}

\subsection{Dataset Generation}
\label{sec:dataset}
We utilize a Xilinx UltraScale+ \blue{XCZU9EG} FPGA contained on the ZCU102 board. 
We generate multiple bitstreams to configure the FPGA board with tenant designs. To replicate a multi-tenancy scenario, we employed $20$ distinct reconfigurable regions within the FPGA. These regions \blue{vary} in size, ranging from $50\%$ of the FPGA resources (with room for an additional equally-sized tenant) to $15\%$ of the FPGA resources (allowing space for up to five other equally-sized tenants).

Based on these tenant regions, we built a dataset of $115$ netlists.
The netlists are based on benign designs from the ISCAS, Groundhog, Berkley, and OpenCores benchmarks in addition to a benign AES implementation to compare how well it can be differentiated from the maliciously modified AES.
The netlists contain malicious designs as well, such as the MUX-based and Latch-based ROs as explained in~\cite{la2020fpgadefender}, and \ac{AES}-based attacks detailed in~\cite{Provelengios2020}.
Moreover, we build attacks based on \ac{DES} and \ac{SHA} in a method similar to the \ac{AES}-based attacks.
For the \ac{AES}-based attacks, we hide them with circuits from the ISCAS benchmark as explained in~\cite{learnmal23} to check how hard \blue{it is to} detect them.

\begin{table}[tb]
\centering
\caption{The Dataset Generated for Evaluating MaliGNNoma}
\label{tab:dataset}
\resizebox{0.49\textwidth}{!}
{%
\begin{tabular}{cccc}
\hline
 & Design & \# Netlists\\
\hline
\multirow{6}{*}{Malicious} & Modified AES~\cite{Provelengios2020} & 8 \DrawBar{8}\\ 
 & Modified DES* & 14 \DrawBar{14}\\
 & Modified SHA* & 9 \DrawBar{9}\\
 & Hidden attacks~\cite{learnmal23} & 7 \DrawBar{7}\\
 & Latch-RO~\cite{la2020fpgadefender} & 9 \DrawBar{9}\\
 & MUX-RO~\cite{la2020fpgadefender} & 21 \DrawBar{21}\\
\hline
\multirow{5}{*}{Benign} & ISCAS~\cite{iscas89} & 9 \DrawBar{9}\\
 & Groundhog~\cite{Groundhog10} & 10 \DrawBar{10}\\
 & Berkley~\cite{2007benchmarking} & 4 \DrawBar{4}\\
 & OpenCores~\cite{OpenCores} & 16 \DrawBar{16}\\
 & AES~\cite{seqAES} & 8 \DrawBar{8}\\
\hline
* Indicates Own design & & \\
\end{tabular}

}
\end{table}

Table~\ref{tab:dataset} shows the designs used and their numbers; overall we have 68 malicious designs and 47 benign designs.
It should be noted that for the benign modules, it is not the same design repeated several times but rather several designs from the same benchmark, e.g., artificial neural network and bitcoin miner from OpenCores.
It is worth noting that not all designs were compatible with every reconfigurable region due to resource limitations. Therefore, some designs were omitted from specific tenant regions if they could not fit within the assigned resources.
This is not unreasonable as in real multi-tenant scenarios some tenant designs would not be able to use the smaller cheaper tenants regions.

Through these dataset generation techniques, we aimed to capture a wide range of realistic scenarios, considering resource constraints, design modifications, and potential attack concealment, enabling comprehensive evaluation of our GNN-based malicious netlist classifier for secure cloud FPGAs.

\subsection{MaliGNNoma Integration}
The MaliGNNoma pipeline may initially appear independent of the FPGA netlist scanning task. However, due to the trainable weights involved in the message passing step of the GNN, the MaliGNNoma procedure becomes optimized through training on the dataset comprising malicious and benign modules. Consequently, MaliGNNoma becomes specifically tailored to effectively differentiate between malicious and benign modules by mapping the graphs in an optimal manner.
\begin{figure}[!t]
\centering
\includegraphics[width=0.3\textwidth]{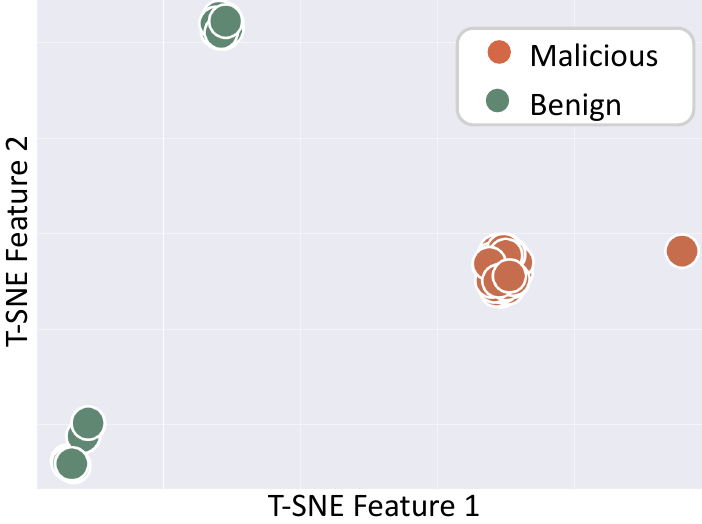}
\caption{MaliGNNoma graph embeddings for benign and malicious samples. The figure visually represents the 2-D T-SNE projection of the generated embeddings by MaliGNNoma for each graph. This visualization illustrates the optimized embedding generation, specifically tailored to separate malicious from benign circuits.}

\label{fig:GNN_tsne}
\end{figure}
\begin{figure*}[!t]
\centering
\includegraphics[width=0.98\textwidth]{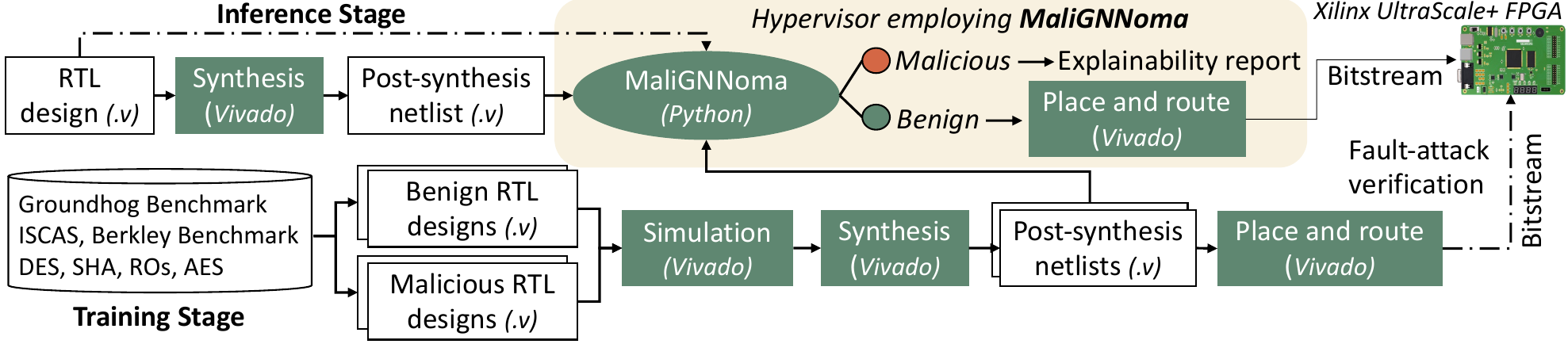}

\caption{MaliGNNoma experimental setup and tool flow.}
\label{fig:exp_flow}
\end{figure*}
\blue{To visually demonstrate the effectiveness of MaliGNNoma in extracting meaningful representations from netlists, we utilize t-distributed stochastic neighbor embedding (t-SNE)~\cite{maaten2008visualizing} to map the GNN-generated embeddings of the dataset samples (discussed in Section~\ref{sec:dataset}) onto a two-dimensional space.}

\blue{The resulting visualization in Fig.~\ref{fig:GNN_tsne} shows distinct clusters for malicious and benign samples, with no overlap, allowing for accurate classification. This highlights the optimization of the GNN stage to enhance the separation between malicious and benign FPGA netlists.}

\section{Evaluation}
We summarize the experimental setup and the process of MaliGNNoma training in Fig~\ref{fig:exp_flow}. Next, we describe the experimental setup in detail.
\subsection{Experimental Setup}
\subsubsection{Evaluation Metrics}
We report the classification Accuracy, Precision, Recall, and F1 score, calculated as follows 
\begin{equation}
\footnotesize
Prec. = \frac{TP}{TP+FP}, Recall = \frac{TP}{TP+FN}, F_1 = \frac{2 \times Prec. \times Recall}{Prec.+Recall}
\end{equation}

$TP$ and $FP$ denote true positives and false positives. Accuracy measures the percentage of correctly classified samples.

For evaluating the GNN explanation, we utilize the fidelity- metric. Fidelity assesses the impact of the generated explanatory subgraph $G_S$ on the original prediction $\hat{y}$. This can be evaluated by providing only the subgraph to the model (fidelity-). Fidelity measures the extent to which an explainable model accurately replicates the underlying natural phenomenon or the logic of the GNN model. The metric is calculated as follows, where $D$ indicates the number of testing samples, $y_i$ indicates the ground truth for sample $i$, and $\hat{y}_i^{G_S}$ indicates the model's prediction for $G_S$:

\begin{equation}
  \textrm{fid}_{-} = \frac{1}{D} \sum_{i = 1}^D \left\| \mathbb{1}(\hat{y}_i = y_i) - \mathbb{1}( \hat{y}_i^{G_S} = y_i) \right\|
\end{equation}
$\mathbb{1}$ represents the \texttt{indicator} function that takes on the value $1$ when a certain condition is true and $0$ when the condition is false. A sufficient explanation has a fidelity- score close to 0.

\subsubsection{GNN Parameters}
Since GNNs have not been previously applied to FPGA netlist analysis, in our experimental investigation, we explore two different GNN baselines, specifically the GIN and GCN, across various GNN configurations to assess their effectiveness.

For the GCN, we consider a baseline GCN with 2 to 3 layers and hidden dimensions of 100, 200, and 400. We also experiment with different \textit{dropout} rates in the range of 0.1, 0.2, and 0.5,\footnote{Dropout is a regularization technique used to prevent overfitting in neural networks by randomly dropping out (setting to zero) some of the units in a layer during training.} after every GNN layer, along with the addition of an attention pooling layer with pooling ratios of 20\%, 50\%, and 80\%. Additionally, we explore different \texttt{READOUT} functions, including \texttt{Max}, \texttt{Mean}, and \texttt{Sum}.

For the GIN, we consider 2 and 3 layers with hidden dimensions of 64, 100, and 128. The default setup for the GIN employs a \texttt{Sum} \texttt{READOUT} function without a pooling layer. We add a dropout stage after the linear layer.

All models were trained for 100 epochs using the mini-batch gradient descent algorithm with a batch size of 4 and a learning rate of 0.001.

\subsubsection{Validation}
For each sample in the dataset, a leave-one-out cross-validation approach is applied. The dataset is randomly shuffled, and for each sample, it is divided into a training set and a validation set (with a 10\% split of the remaining data). During training, the GNN is evaluated on the validation set every 10 epochs. The GNN parameters that yield the best performance on the validation set are selected as the final model for testing. This process is carried out once for each sample in the dataset, with one sample left out for testing in each iteration.

\subsection{Results}

\subsubsection{Classification Performance} The best MaliGNNoma performance was achieved using 2 layers of GIN with a hidden dimension of 100 and a dropout rate of 0.5. This resulted in an accuracy of 98.24\%, precision of 97.88\%, recall of 98.5\%, and an F1 score of 98.19\%. MaliGNNoma effectively detects malicious netlist configurations, even sophisticated ones like AES-based, and achieves state-of-the-art performance in all aspects. Table~\ref{tab:dataset_result} presents the classification performance of MaliGNNoma on the various subsets of the dataset. As evident from the table, there were only two misclassifications in the entire dataset: one benign AES was misclassified as malicious, and one Latch-RO circuit was misclassified as benign.

\begin{table}[tb]
\centering
\caption{MaliGNNoma Classification Results}
\label{tab:dataset_result}
\resizebox{0.49\textwidth}{!}
{%
\begin{tabular}{cccc}
\hline
 & Design & \# Netlists Correctly Classified/Total \#\\
\hline
\multirow{6}{*}{Malicious} & Modified AES~\cite{Provelengios2020} & 8/8\\ 
 & Modified DES & 14/14 \\
 & Modified SHA & 9/9\\
 & Hidden attacks~\cite{learnmal23} & 7/7 \\
 & Latch-RO~\cite{la2020fpgadefender} & 8/9\\
 & MUX-RO~\cite{la2020fpgadefender} & 21/21\\
\hline
\multirow{5}{*}{Benign} & ISCAS~\cite{iscas89} & 9/9\\
 & Groundhog~\cite{Groundhog10} & 10/10\\
 & Berkley~\cite{2007benchmarking} & 4/4\\
 & OpenCores~\cite{OpenCores} & 16/16\\
 & AES~\cite{seqAES} & 7/8\\
\hline

\end{tabular}

}
\end{table}

\subsubsection{Comparison of GNN Architectures}
Comparing GCN to GIN, with the same hidden dimension of 100, GCN (with \texttt{Max} \texttt{READOUT}) achieves similar performance, with an accuracy of 94.73\%. However, to achieve this result, a pooling ratio of 20\% was applied, which was unnecessary for GIN. Overall, when comparing both, GIN exhibits greater expressiveness as it incorporates a neural network in the aggregation step, enabling it to capture more intricate relationships in the graph data. However, for both architectures, increasing the number of layers to 3 leads to a significant reduction in performance. For instance, in the case of GIN, this increase resulted in a notable accuracy drop to 77.19\%. This effect is commonly referred to as ``over-smoothing.'' Over-smoothing occurs when nodes lose their discriminative power because they aggregate information from too many neighbors, making them nearly identical.

\subsubsection{GNN Explanation}
After model training and testing, the model is passed to the PGExplainer to extract an explanatory subgraph from the entire graph representation of the netlist. This step is particularly crucial, especially in cases where a malicious label is predicted, as it provides justification for the model's decision.
To evaluate the performance of subgraph extraction without relying on node labeling to identify malicious components, we use the fidelity- metric. An explanation is considered sufficient if it independently leads to the initial prediction of the model. Since there may be other configurations in the graph that could also lead to the same prediction, it is possible to have multiple sufficient explanations for a single prediction. A sufficient explanation is characterized by a fidelity- score close to 0.
For our dataset, the average fidelity- score obtained is 0.28, with a fidelity- score of 0 being observed for 70\% of the samples. This indicates that in the majority of cases, the extracted explanations are effective in justifying the model's predictions.\footnote{The output of the GNN explainer represents connectivity (edges) between circuit components, not a list of primitives. We align with the state-of-the-art in ML, reporting explainability performance through the fidelity metric. This metric assesses whether removing the detected subgraph from the original graph and retesting the model can alter predictions, which we believe is a meaningful evaluation. Our findings reveal that the reported nodes/edges may not necessarily form a cluster within the circuit, especially in AES-based attacks where the entire circuit contributes to the fault-injection attack.}

\subsubsection{Runtime}
MaliGNNoma runs on a single machine of $10$ cores ($2$x Intel(R) Xeon(R) CPU E5-2680 v4@2.4GHz). On average, MaliGNNoma training process for $100$ epochs, including dataset reading and conversion to graphs, takes approximately $6$ hours. It is important to note that this is an offline procedure. Once MaliGNNoma is trained, the inference time for each prediction averages around $0.05$ seconds.

\subsubsection{Comparison to Related Works}

As we show in Table~\ref{tab:comparison}, not all attack vectors are detected by the state-of-the-art.
To quantify this, we compare MaliGNNoma against them based on our dataset. Table~\ref{tab:acc_comparison} shows the results of this comparison. 
As these tools are not openly available, we perform a conservative comparison, i.e., we assume they have 100\% accuracy in detecting the attacks they report capable of detecting.
This is to their benefit, as realistically they have lower accuracy e.g., as reported in~\cite{learnmal23,CNNbased,chaudhuri2023diagnosis}.
Even with this conservative comparison, MaliGNNoma outperforms all the state-of-the-art solutions.
The highest accuracy is theoretically reached by the tool of~\cite{learnmal23} which is still lower than the actual accuracy MaliGNNoma reaches.

\begin{table}[tb]
\centering
\caption{\textsc{Comparison to Related Work based on the Type of Attacks they Mention they are Capable of Detecting. The comparison is conservative, giving them 100\% accuracy in detecting the attacks mentioned in their works.}}
\label{tab:acc_comparison}
\resizebox{0.49\textwidth}{!}{%
\begin{tabular}{ccccccc}
\hline
 & \textbf{\begin{tabular}[c]{@{}c@{}}Tool\\ of~\cite{learnmal23}\end{tabular}} & \textbf{\begin{tabular}[c]{@{}c@{}}Tool\\ of~\cite{CNNbased}\end{tabular}} & \textbf{\begin{tabular}[c]{@{}c@{}}Tool\\ of~\cite{chaudhuri2023diagnosis}\end{tabular}} & \textbf{\begin{tabular}[c]{@{}c@{}}Tool\\ of~\cite{la2020fpgadefender}\end{tabular}} & \textbf{\begin{tabular}[c]{@{}c@{}}Tool\\ of~\cite{Krautter2019mitigating}\end{tabular}} & \textbf{\begin{tabular}[c]{@{}c@{}}MaliGNN-\\ oma\end{tabular}} \\\hline
\textbf{Accuracy} & 80\% & 66.9\% & 86.1\% & 66.9\% & 66.9\% & \textbf{98.24\%} 
\\ \hline
\end{tabular}
}
\end{table}
\subsection{Discussion}
\subsubsection{Netlist Versus Bitstream}
There may be concerns about processing the netlist instead of the bitstream due to potential piracy issues. However, it is important to note that a bitstream is not inherently secure and can be reverse-engineered~\cite{bitman}. Providing the netlist instead of the bitstream does not introduce any additional threats in terms of security. Only encrypted bitstreams can provide security, but they cannot be analyzed since they are indistinguishable from random data. The topic of incorporating obfuscation mechanisms in the netlist to make reverse engineering more difficult is a separate research direction that is not directly related to our work.
Alternative to obfuscation, to preserve the security of the bitstreams, \cite{zeitouni21} shows an attestation scheme including bitstream checking for cloud FPGAs through a trusted third party.
This attestation scheme or similar FPGA attestation schemes can be used for the netlist checking for malicious attacks, but their implementation is out of the scope of this work.

\subsubsection{Security of GNNs}
Recent research highlights the vulnerability of GNNs themselves to \textit{poisoning} and \textit{backdoor} attacks, particularly when training is outsourced~\cite{alrahis2023poisonedgnn,alrahis2023graph}. To mitigate this threat, there is a growing focus on developing robust GNN implementations. MaliGNNoma offers explainability and the potential to detect backdoor attacks. However, the specific application of explainability mechanisms for detecting backdoor attacks in GNNs analyzing circuits has not yet been extensively explored.
\section{Conclusions}
\bluee{We introduce MaliGNNoma, a defense method based on graph neural networks (GNNs) designed to ensure the security of cloud field-programmable gate arrays (FPGAs), including, but not limited to, multi-tenancy scenarios. MaliGNNoma can be employed by a hypervisor to scan input netlists for potential malicious circuit implementations before FPGA configuration, serving as the initial security layer in a much-needed multi-tiered security system. Unlike other scanning methods, MaliGNNoma directly learns from netlists instead of bitstreams, enabling it to detect a wide range of attacks with high accuracy.} Our work showcases the first application of GNNs for FPGA security. Further, we open-source MaliGNNoma.

\section*{Acknowledgments}
This work was partially funded by the ``Helmholtz Pilot Program for Core Informatics (KiKIT)'' at Karlsruhe Institute of Technology (KIT), and the German Federal Ministry of Education and Research (BMBF) through the Software Campus Project. Further, the work of Lilas Alrahis was supported by the KIT International Excellence Fellowship.
\bibliographystyle{IEEEtran}
\bibliography{main.bib}

\end{document}